\documentstyle[sprocl]{article}
\begin{document}

\hfill\vbox{
  \hbox{OHSTPY-HEP-T-96-021}
  \hbox{hep-ph/9608442}
  \hbox{August 1996}
}\bigskip

\title{THERMODYNAMICS OF QCD AT HIGH
TEMPERATURE~\footnote{~Invited talk presented at
DPF96 in Minneapolis, August 1996.}
}

\author{ A. NIETO }

\address{Department of Physics,
The Ohio State University, Columbus, OH 43210, USA}
\maketitle\abstracts{
A hierarchy of effective field theories is used to
separate the contributions from different momentum scales
and to calculate the free energy of QCD at high temperature
in powers of the coupling constant up to order $g^5$.
The behavior of the perturbative series will also be
discussed.
}
The partition function of QCD at finite temperature is
obtained from the Lagrangian $ {\cal L}_{\rm QCD} = (1/ 4)
G_{\mu\nu}^a G_{\mu\nu}^a + \overline{q}\gamma_\mu D_{\mu}
q$, with gauge coupling constant $g$. We consider the
effective theory that results from integrating out the
nonstatic modes. Such a theory, which is called
Electrostatic QCD (EQCD), is made out of only the static
bosonic modes and 3 dimensional.  The free energy density is
$ F_{\rm QCD} = T \left( f_E - \log {\cal Z}_{\rm EQCD}/ V
\right) $, where ${\cal Z}_{\rm EQCD}$ is the partition
function of EQCD obtained from the Lagrangian
\begin{equation}
  {\cal L}_{\rm EQCD} = {1\over 4} G_{ij}^a G_{ij}^a +
    {1\over 2} (D_i A_0)^2 + {1\over 2} m_E^2 A_0^2 +
    {1\over 8} \lambda A_0^4 + \delta{\cal L}_{\rm EQCD}
\end{equation}
whith effective gauge coupling constant $g_E$.  $\delta{\cal
L}_{\rm EQCD}$ represents an infinite series of
non-renormalizable terms. The effects of the fermions are
incorporated into the effective parameters.  By power
counting, the effective parameters at leading order are:
$f_E\sim T^3$, $m_E^2\sim g^2 T^2$, $g_E^2\sim g^2 T$, and
$\lambda_E\sim g^4 T$.

The magnetostatic fields remain massless; therefore, we can
go further in separating the different scales of QCD at high
temperature by integrating out $A_0$. We obtain an effective
theory of EQCD which is called magnetostatic QCD (MQCD). The
free energy density of QCD can be written as $F = T ( f_E + f_M
+f_G)$, where $f_G = - \log {\cal Z}_{\rm MQCD}/ V$, and
${\cal Z}_{\rm MQCD}$ is the partition function of MQCD
obtained from the Lagrangian $ {\cal L}_{\rm MQCD} = (1/4)
G_{ij}^a G_{ij}^a + \delta{\cal L}_{\rm MQCD}\, .  $ The
gauge coupling constant is $g_M$ and $\delta{\cal L}_{\rm
MQCD}$ represents an infinite series of non-renormalizable
terms.

Again, we can use power counting to identify the order of
the leading contribution to the MQCD parameters: $f_M\sim
m_E^3\sim g^3 T^3$ and $g_M^2\sim g_E\sim g^2 T$.
$\lambda_E$ contributes to the free energy only at order
$g^6$; so, if we are interested in the free energy at lower
order, we can ignore $\lambda_E$. Similarly the
non-renormalizable terms of EQCD can also be omitted.

The leading contribution to $f_G$ can be obtained by
realizing that the only parameter with dimensions involved
in MQCD is $g_M$ and therefore the leading contribution is
of order $g_M^6\sim g^6 T^3$. Since we are interested in
computing the free energy of QCD up to order $g^5$ we can
ignore the contribution from $f_G$; it will be enough to
compute $f_E$ and $f_M$. We conclude that the free energy of
QCD up to order $g^5$ is given by $F = T[ f_E(T,g;\Lambda_E)
+ f_M(m^2_E, g_E;\Lambda_E)]$, where $\Lambda_E$ is a
factorization scale that separates the scales $T$ and $gT$.
Therefore, we have to determine $f_E$, $m_E^2$, $g_E$, and
$f_M$. Calculations for $SU(N)$ gauge theory with $n_f$
fermions and results in analytical form are detailed
in~\cite{bn5}; here, I will just state the results in
numerical form for $SU(3)$ with 3 fermions.

The effective mass $m_E$ is computed by matching the
electrostatic screening mass for QCD and EQCD.
\begin{equation}
m_E^2 =  0.93\; {g^2(\mu)\over 4\pi}\; T^2
\left[ 1 +
\left( -\; 0.26 + 1.43\; \log {\mu \over 2 \pi T} \right)
         {g^2 \over 4\pi}
\right] \, .
\label{mE}
\end{equation}
At this order in $g^2$, there is no dependence on the
factorization scale $\Lambda_E$.

The parameter $f_E$ is determined by calculating the
free energy in both full QCD and EQCD, and matching the two
results.
\begin{eqnarray}
  f_E(\Lambda_E) &=& - 14.8\; T^3\; \left[
    1 -\; 0.90\; {g^2 \over 4\pi} \; + \right.
\nonumber \\
  &+& \left. \left(
    18.3 + 6.91\;\log{\Lambda_E \over 2 \pi T} -
    1.30\;\log{\mu \over 2 \pi T} \right)
      \left( {g^2 \over 4\pi} \right)^2
\right] \, ,
\label{fE}
\end{eqnarray}
where $g$ is the coupling constant in the $\overline{\rm
MS}$ renormalization scheme at the scale $\mu$.  We have
used the renormalization group equation of the coupling
constant to shift the scale of the running coupling constant
to an arbitrary renormalization scale $\mu$.

Through order $g^5$, $f_M$ is proportional to the logarithm
of the partition function for EQCD: $f_M \;=\; - \log {\cal
Z}_{\rm EQCD}/ V $.  Now, we have to consider the
cotnribution to $\log {\cal Z}_{\rm EQCD}$ of orders $g^3$,
$g^4$, and $g^5$ which are given by the sum of 1-loop,
2-loop, and 3-loop diagrams.  The details of this
calculation can be found in~\cite{bn5}; the final result is
\begin{equation}
  f_M(\Lambda_E) = - 0.21 m_E^3 \left[
    1 - \left(0.54 + 0.72 \log{\Lambda_E\over 2m_E}\right)
      \left( {g_E^2 \over m_E}\right) -
    0.70 \left( {g_E^2 \over m_E}\right)^2
\right] \, .
\label{fM}
\end{equation}

The coefficient $f_M$ in (\ref{fM}) can be expanded in
powers of $g$ by setting $g_E^2 = g^2T$ and by substituting
the expression (\ref{mE}) for $m_E^2$.  The complete free
energy to order $g^5$ is then $F = (f_E + f_M) T$; in
agreement with the result obtained independently by
Kastening and Zhai~\cite{kastening-zhai}. Note that the
dependence on the arbitrary factorization scale $\Lambda_E$
cancels between $f_E$ and $f_M$, up to corrections that are
higher order in $g$, leaving a logarithm of $T/m_E$.

We have calculated the free energy as a perturbation
expansion in powers of $g$ to order $g^5$. We now ask how
small $\alpha_s\equiv g^2/(4\pi)$ must be in order for the
perturbation expansion to be well-behaved~\cite{bn5,bn4}. If
the series is apparently convergent, then it can plausibly
be used to evaluate the free energy.  For simplicity, we
consider the case $n_f=3$ and choose the renormalization
scale $\mu = 2 \pi T$ which is the mass of the lightest
nonstatic mode. The correction to the leading order term of
the free energy is a multiplicative factor $1 - 0.9 \alpha_s
+ 3.3 \alpha_s^{3/2} + (7.1 + 3.5
\log\alpha_s)\alpha_s^2 - 20.8\alpha_s^{5/2}$. If the
temperature is a few times above the transition temperature
($\sim 200$ MeV); {\em e.g.\/}, $T\simeq 350$ MeV the series
has the form $1 - 0.27 + 0.54 + 0.26 - 1.03 + \cdots$. It is
clear that the perturbative series is not convergent at {\em
this\/} temperature.

$f_E$ and $m_E$ contain the contributions of order $T$.  In
the case of $m_E^2$, which is given by~(\ref{mE}), the
correction to the leading order result form the series $1 -
0.26\alpha_s$.  The next-to-leading-order correction to
$m_E^2$ is smaller than the leading-order term only if
$T>40$ MeV. A similar analysis of $f_E$ gives $T>67$ MeV.
We conclude that the perturbation series for the parameters
of EQCD are well-behaved provided that $T>70$ MeV.

Expression~(\ref{fM}) gives $f_M$ which is the contribution
from the scale $T$.  The next-to-leading-order correction is
small if $\Lambda_E$ is chosen to be approximately $m_E$.
The next-to-next-to-leading-order correction is smaller than
the leading order term only if $T>2$ GeV.  Thus the
perturbation series for $f_M$ is well-behaved only for
temperatures that are much higher than those required for
the parameters of EQCD to have well-behaved perturbation
series.

This analysis indicates that the slow convergence of the
expansion for $F$ in powers of $\sqrt{\alpha_s}$ can be
attributed to the slow convergence of perturbation theory at
the scale $gT$.

\section*{Acknowledgments}
This work was supported in part by the U.S. Department of
Energy, Division of High Energy Physics, under Grant
DE-FG02-91-ER40690. I would like to thank E.~Braaten for
valuable discussions.

\end{document}